\documentclass[aps,superscriptaddress,twocolumn]{revtex4-2}

\usepackage{epsfig}
\usepackage{latexsym}
\usepackage{graphicx}
\usepackage{amssymb}
\usepackage{multirow}
\usepackage{tabularx}
\usepackage{bm}
\usepackage{tikz}
\usepackage{latexsym}
\usepackage{bm}
\usepackage{framed}
\usepackage{color}
\usepackage{amsmath}
\usepackage{hyperref}

\begin{document}

\title{Designing metachronal waves of cilia}

\author{Fanlong Meng}
\thanks{F. M. and R. R. B. contributed equally to this work. }
\affiliation{CAS Key Laboratory for Theoretical Physics, Institute of Theoretical Physics, Chinese Academy of Sciences, Beijing 100190, China}
\affiliation{Max Planck Institute for Dynamics and Self-Organization (MPIDS), G\"ottingen 37077, Germany}

\author{Rachel R. Bennett}
\thanks{F. M. and R. R. B. contributed equally to this work. }
\affiliation{School of Mathematics, University of Bristol, Bristol BS8 1TW, UK}
\affiliation{Rudolf Peierls Centre for Theoretical Physics, University of Oxford, Oxford OX1 3NP, UK}

\author{Nariya Uchida}
\affiliation{Department of Physics, Tohoku University, Sendai, 980-8578, Japan}

\author{Ramin Golestanian}
\email{ramin.golestanian@ds.mpg.de}
\affiliation{Max Planck Institute for Dynamics and Self-Organization (MPIDS), G\"ottingen 37077, Germany}
\affiliation{Rudolf Peierls Centre for Theoretical Physics, University of Oxford, Oxford OX1 3NP, UK}

\date{\today}

\begin{abstract}
On surfaces with many motile cilia, beats of the individual cilia coordinate to form metachronal waves. We present a theoretical framework that connects the dynamics of an individual cilium to the collective dynamics of a ciliary carpet via systematic coarse-graining. We uncover the criteria that control the selection of frequency and wavevector of stable metchacronal waves of the cilia and examine how they depend on the geometric and dynamical characteristics of single cilia, as well as the geometric properties of the array. We perform agent-based numerical simulations of arrays of cilia with hydrodynamic interactions and find quantitative agreement with the predictions of the analytical framework. Our work sheds light on the question of how the collective properties of beating cilia can be determined using information about the individual units, and as such exemplifies a bottom-up study of a rich active matter system.
\end{abstract}

\maketitle

Motile cilia are hair-like organelles that beat with a whip-like stroke that breaks time-reversal symmetry to create fluid flow or propel swimming microorganisms under low Reynolds number conditions \cite{Gray:1928,Brennen:1977,Golestanian2011}. The beat is actuated by many dynein motors, which generate forces between microtubules that cause the cilium to bend in a robust cyclic manner with moderate fluctuations~\cite{Camalet:1999,Friedrich2014}. On surfaces with many cilia, the actuating organelles can coordinate with each other and collectively beat in the form of a metachronal waves, where neighboring cilia beat sequentially (i.e. with a phase lag) rather than synchronously \cite{Knight-Jones1954}.
The flows created from this coordinated beating are important for breaking symmetry in embryonic development \cite{Nonaka:2002,Takamatsu2013}, creation of complex and dynamic flow patterns for the cerebrospinal fluid in the brain \cite{Faubel176,Pellicciotta8315}, and providing access to nutrients \cite{Short:2006}. In microorganisms such as {\it Paramecium} and {\it Volvox}, the metachronal beating of cilia provides propulsion strategies in viscous environments \cite{Tamm:1975,Brumley2015}. It has been shown that depending on the parameters beating ciliary carpets can exhibit globally ordered and turbulent flow patterns \cite{Uchida2010}, which can be stable even with a moderate amount of quenched disorder \cite{Uchida2010b}, and that metachronal coordination optimizes the efficiency of fluid pumping \cite{Osterman2011,Elgeti2013}. Natural cilia have inspired various designs of artificial cilia \cite{Evans2007, Vilfan2010, Coq2011, Sanchez2011, meng2019}, which may be used for pumping fluid \cite{Gauger2009, Khaderi2011a} and mixing \cite{Matsunaga2019}, or fabrication of microswimmers \cite{Dreyfus2005}.

Hydrodynamic interactions have been shown to play a key role in coordinated beating of cilia \cite{Guirao2007,Brumley2014} and mediating cell polarity control \cite{Guirao2010}. To achieve synchronization between two cilia via hydrodynamic interactions, it is necessary to break the permutation symmetry between them, e.g. by exploiting the dependence of the drag coefficient on the distance from a surface \cite{Vilfan2006}, flexibility of the anchoring of the cilia \cite{Qian2009}, non-uniform beat patterns \cite{Uchida2011, Uchida2012}, or any combination of these effects \cite{Maestro2018}. In addition to the hydrodynamic interactions, the basal coupling between cilia can also facilitate the coordination \cite{Narematsu:2015,Wan2016, Liu2018}.

How can we predict the collective behavior of arrays of many cilia coordinated by hydrodynamic interactions, and in particular the properties of the emerging metachronal waves, from the single-cilium characteristics? Extensive numerical simulations using explicitly resolved beating filaments \cite{Gueron1997, Kim2006, Guirao2007, Osterman2011, Elgeti2013, Ding2014} and simplified spherical rotors \cite{Uchida2010, Wollin2011, Brumley2015, Ghorbani2017} have demonstrated that metachronal coordination emerges from hydrodynamic interactions. However, insight into this complex many-body dynamical system at the level that has been achieved in studies of two cilia is still lacking. Here, we propose a theoretical framework for understanding the physical conditions for coordination of many independently beating cilia, which are arranged on a substrate in the form of a 2D array immersed in a 3D fluid. We uncover the physical conditions for the emergence of stable metachronal waves, and predict the properties of the wave in terms of single-cilium geometric and dynamic characteristics.

\begin{figure}[t]
\centering
\includegraphics[width=0.78\columnwidth]{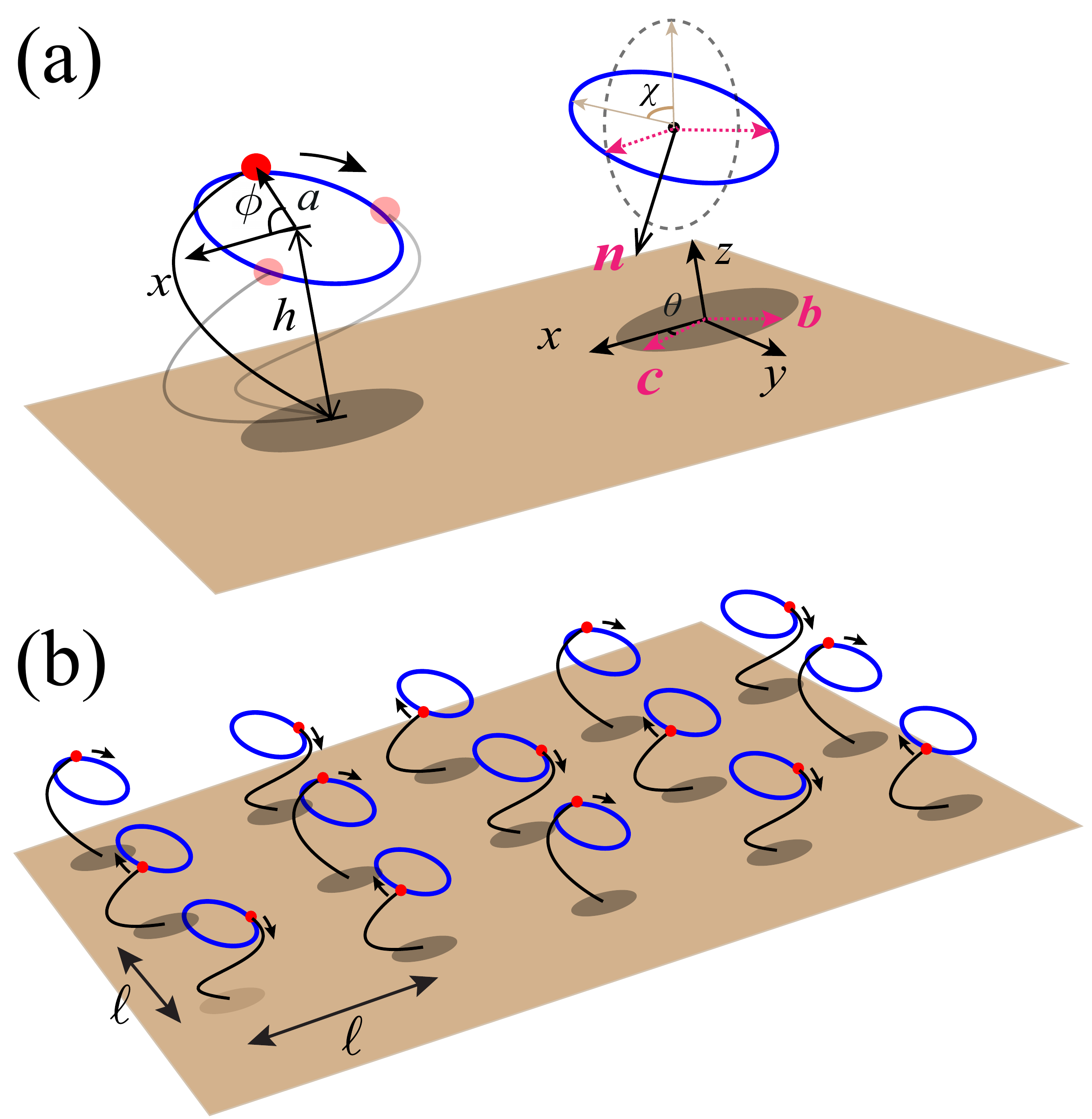}
\caption{(a) Simplified description of a single cilium as represented by a force-monopole with a cyclic trajectory above a rigid substrate. Various geometric measures and orientation vectors are defined in the figure. (b) An array of cilia on a square lattice in the $x$-$y$ plane with the lattice constant $\ell$.}
\label{setup}
\end{figure}

\section*{Results}

We use a simplified model of a cilium as a force monopole 
moving along a circular trajectory of radius $a$ above a substrate
, as shown in Fig.~\ref{setup}(a). To theoretically study metachronal coordination, we consider such model cilia on a lattice of spacing $\ell$ in the $x$-$y$ plane as shown in Fig.~\ref{setup}(b). 
We examine the role of the geometric parameters in determining the collective mode of coordination. We parametrize the orientation of the cilia by the angle $\theta$ that the plane of the circular trajectory makes with the $\bm{e}_{x}$ direction, and the tilt angle $\chi$ it makes with the $\bm{e}_{z}$ direction. More concretely, we define the vector $\bm{c} = (\cos\theta,\sin\theta, 0)$ to characterize the 2D orientation of the circular orbit and $\bm{b} = (-\sin\theta, \cos\theta, 0)$, which is perpendicular to $\bm{c}$, and the unit vector that is normal to circular trajectory is defined as $\bm{n} = -\cos\chi \bm{b}  - \sin\chi \bm{e}_{z}$; see Fig.~\ref{setup}(a). The position of the sphere representing the force monopole along the trajectory, which we parametrize by the polar angle $\phi_i$ for the \emph{i}th sphere, is 
\begin{equation}
\bm{R}_{i}=\bm{r}_{i}+h \bm{e}_{z}+a \cos\phi_i \bm{c}+a\sin\phi_i ( -\sin\chi \bm{b}  + \cos\chi \bm{e}_{z}), 
\end{equation}
where $\bm{r}_{i}=(x_{i},y_{i},0)$ denotes the lattice coordinate (with spacing $\ell$) and $h$ denotes the distance from the center of the trajectory to the substrate.
There are also simulation studies adopting non-circular trajectories of the cilia, and interested readers can refer to Ref. \cite{Vilfan2006, Brumley2012, Brumley2015, Ghorbani2017}.

\subsection*{Dynamical equations}
Each cilium is driven independently by a tangential force acting on the bead. The magnitude of the force, $f(\phi_{i})$, depends only on the location of the bead along its trajectory, and the direction of the force is given by the tangent vector, which is defined as 
\begin{equation}
\bm{t}_{i}(\phi_{i})=\frac{d \bm{R}_{i}/d \phi_i}{|d \bm{R}_{i}/d \phi_i|}.
\end{equation}
The friction coefficient of the bead, $\zeta(\phi_i)$, can, in general, depend on the location of the bead along the trajectory, e.g. due to the proximity of a substrate. Balancing the forces for a single cilium, we find that the velocity of the bead is $\bm{v}_i=\dot{\phi}_i a \,\bm{t}_i= f(\phi_i)/\zeta(\phi_i)\, \bm{t}_i$. In an array, the hydrodynamic interactions between the cilia will also influence the beating cycle, leading to a system of coupled dynamical governing equations for the phase variables
\begin{eqnarray}
\frac{d {\phi}_i}{d t} = \frac{f(\phi_i)}{\zeta(\phi_i) a}+\frac{1}{a}\sum_{j} \bm{t}_i(\phi_i)\cdot \bm{G}(\bm{R}_i,\bm{R}_j)\cdot\bm{t}_j(\phi_j) f(\phi_j),
\label{Mo1}
\end{eqnarray}
where the the Green's function $\bm{G}(\bm{R},\bm{R'})$ (the Blake tensor) represents the hydrodynamic effect in an incompressible fluid of a force-monopole located at $\bm{R'}$ at the observation point $\bm{R}$ in the presence of a substrate with no-slip boundary condition~\cite{Blake1971}.

The intrinsic angular speed of each cilium given as 
\begin{equation}
\Omega(\phi)=f(\phi)/[\zeta(\phi) a], 
\end{equation}
can generically have phase dependence arising from the stroke pattern of the beating, which can be represented via its harmonics as 
\begin{equation}
f(\phi) = f_{0}[1+\sum_{n=1}A_{n}\cos n\phi+B_{n}\sin n\phi], 
\end{equation}
and the cyclic change in the friction, which we represent as 
\begin{equation}
\zeta(\phi)=\zeta_{0}[1+\sum_{n=1}C_{n}\cos n\phi+D_{n}\sin n\phi]. 
\end{equation}
The scale of the friction coefficient can be written as $\zeta_0=4 \pi \eta b$, where the length scale $b$ represents the characteristic (hydrodynamic) size of a cilium. Naturally, the harmonic amplitudes are constrained to values that will correspond to strictly positive values for the force and the friction coefficient. To proceed with the analysis of \eqref{Mo1}, we introduce a coordinate transformation $\phi\rightarrow\bar{\phi}$, defined via the following relation: 
\begin{equation}
\frac{d\bar{\phi}}{d\phi}=\frac{\Omega(\phi)}{\Omega_0},
\end{equation}
where $\Omega_{0}=f_0/(\zeta_0 a)$ is a constant angular speed describing the free dynamics of the new coordinate \cite{Uchida2011}. The definition can be integrated to obtain the relation between the coordinates as
\begin{equation}
\phi(\bar{\phi})\simeq\bar{\phi}+\sum_{n=1}\frac{1}{n}\left[\left(A_{n}-C_n\right)\sin n\bar{\phi}-\left(B_{n}-D_n\right) \cos n\bar{\phi}\right],\\
\label{relationbetweenphase}
\end{equation}
to the lowest order in the harmonic amplitudes.

\begin{figure*}[t]
\centering
\includegraphics[width=2.0\columnwidth]{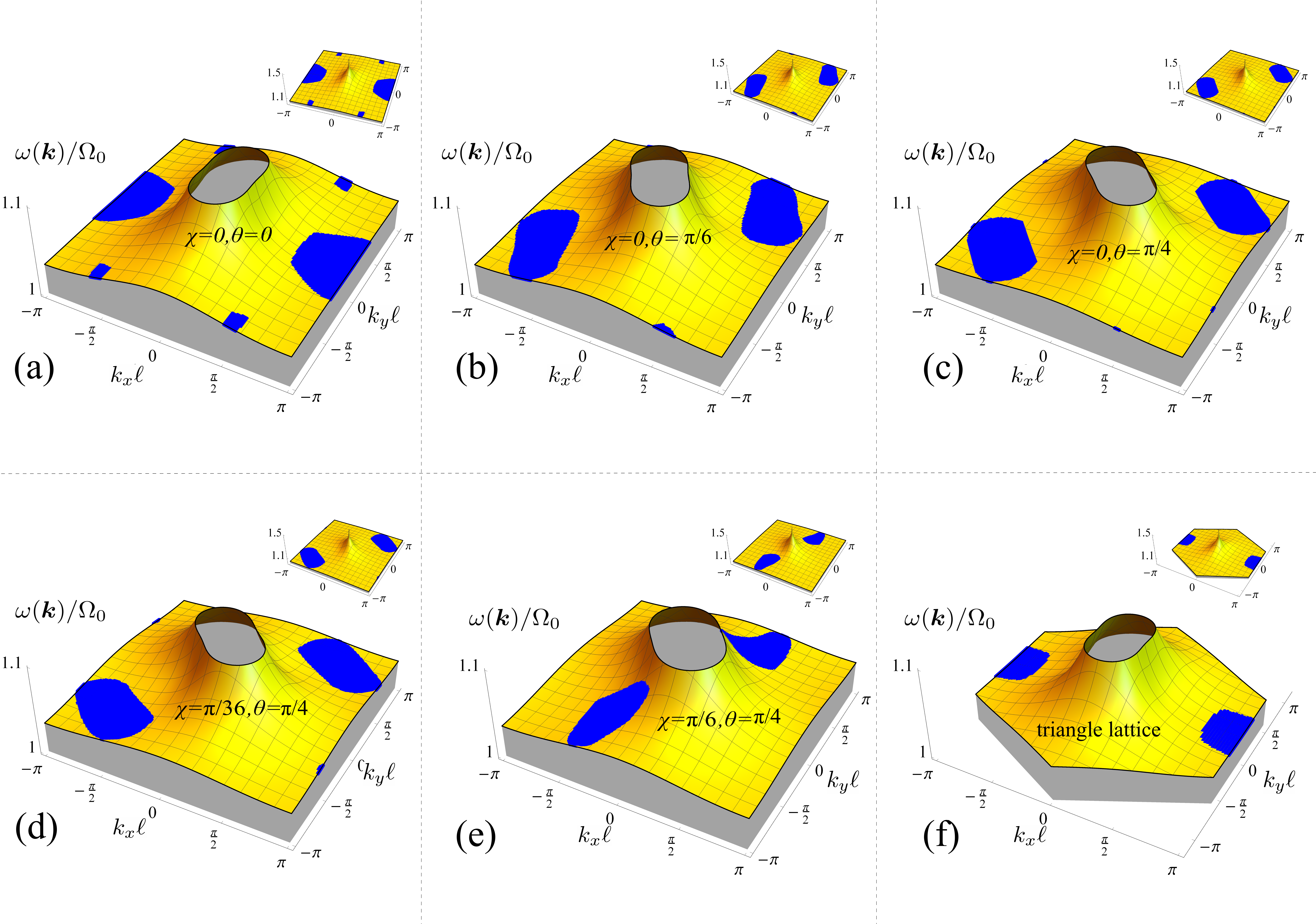}
\caption{Dispersion relation of the metachronal waves. 
The (non-dimensionalized) frequency $\omega(\bm{k})/\Omega_{0}$ defined in Eq.~(\ref{omega5}) is plotted as a function of the wave vector $(k_{x}\ell, k_{y}\ell)$ in the first Brillouin zone. Panels (a)-(e) correspond to a ciliary array on a square lattice with the following trajectory orientation and tilt angles: (a) $\chi=0$, $\theta=0$, (b) $\chi=0$, $\theta=\pi/6$, (c) $\chi=0$, $\theta=\pi/4$, (d) $\chi=\pi/36$, $\theta=\pi/4$ and (e) $\chi=\pi/6$, $\theta=\pi/4$. Panel (f) corresponds to a triangular lattice with $\chi=0$, $\theta=0$. Other parameters for panels (a)-(f) are: $a=0.2 \ell$,  $b=0.05\ell$, $h=\ell$, $A_{2}=0.5$, $B_{2}=0.5$, $C_{2}=0$, and $D_{2}=0$. Blue shaded regions denote the stable wave zones as determined by the linear stability analysis from Eq. (\ref{stability3}). }
\label{relation}
\end{figure*}

Using the translational invariance along the substrate, we can express the Blake tensor in the 2D Fourier space $\bm{q}=(q_{x},q_{y},0)\equiv q \hat{\bm{q}}$ (see Supplemental Material for the details of the derivation) and recast Eq.~(\ref{Mo1}) in terms of the the new coordinate. We then use a separation of time scale between the mean-phase and the phase-difference to simplify the dynamics. By changing the notation from $\phi_i(t)$ to $\phi(\bm{r},t)$ and averaging over the fast variables, the governing dynamical equation can be written as \cite{suppl}
\begin{eqnarray}\label{m2}
&&\partial_t\bar{\phi}(\bm{r}, t)=\Omega_{0}+\frac{\Omega_{0} h b}{16\pi}\,\sum_{\bm{r'}} \int d^2 \bm{q} \,e^{i\bm{q}\cdot(\bm{r}-\bm{r'})} \\
&&\times \big[{\cal M}(\bm{q})\cos\left({\bar{\phi}}(\bm{r})-{\bar{\phi}}(\bm{r}')\right)+{\cal S}(\bm{q})\sin\left({\bar{\phi}}(\bm{r})-{\bar{\phi}}(\bm{r}')\right)\big],\nonumber
\end{eqnarray}
where the $\bm{q}$-dependent kernels are defined as
\begin{eqnarray}
{\cal M}\!\!\!&=&\!\!\!g_0(\bm{q} h)\!+\!g_1(\bm{q} h) (A_{2}-2 C_{2})\!+\!g_2(\bm{q} h) (B_{2}-2D_{2}),\label{M-def}\\
{\cal S}\!\!&=&\!\!g_1(\bm{q} h) (2 B_{2}-D_{2})\!-\!g_2(\bm{q} h) (2 A_{2}-C_{2}),\label{S-def}
\end{eqnarray}
in terms of the following functions: 
\begin{eqnarray}
g_0(\bm{p})\!\!\!\!&=\!\!\!\!&2 p^{-1}[e^{-2p (a/h)}-e^{-2p}][3-\cos^{2}\chi (\hat{\bm{p}}\cdot\bm{c})^{2}] \nonumber \\
\!\!\!\!&+\!\!\!\!& 4(1-p)e^{-2p}[1-\cos^{2}\chi (\hat{\bm{p}}\cdot\bm{b})^{2}] \nonumber \\
\!\!\!\!&-\!\!\!\!& 4(1+p)e^{-2p}\cos^{2}\chi, \nonumber \\
g_1(\bm{p})\!\!\!\!&=\!\!\!\!& p^{-1}[e^{-2p (a/h)}-e^{-2p}][(\hat{\bm{p}}\cdot \bm{b})^{2}-\sin^{2}\chi (\hat{\bm{p}}\cdot\bm{c})^{2}] \nonumber \\
\!\!\!\!&+\!\!\!\!& 2(1-p)e^{-2p}[(\hat{\bm{p}}\cdot \bm{c})^{2}-\sin^{2}\chi (\hat{\bm{p}}\cdot\bm{b})^{2}] \nonumber \\
\!\!\!\!&+\!\!\!\!& 2(1+p)e^{-2p}\cos^{2}\chi,\nonumber \\
g_2(\bm{p})\!\!\!\!&=\!\!\!\!& 2\{p^{-1}[e^{-2p (a/h)}-e^{-2p}]-2(1-p)e^{-2p}\}\sin\chi (\hat{\bm{p}}\cdot \bm{b})(\hat{\bm{p}}\cdot\bm{c}). \nonumber 
\end{eqnarray}
Importantly, we find that only the second harmonics in the beat pattern and the friction cycle play a key role in determining the collective behavior of the cilia at long time scales. The compact form of Eq.~(\ref{m2}) allows us to systematically investigate the conditions under which the array of cilia can admit stable metachronal wave solutions, and what determines the direction of propagation and the wavelength of the wave. As we shall see below, ${\cal M}(\bm{q})$ will determine the characteristics of the metachronal waves and ${\cal S}(\bm{q})$ will determine their stability.

\begin{figure*}[htb]
\begin{center}
\includegraphics[width=2.0\columnwidth]{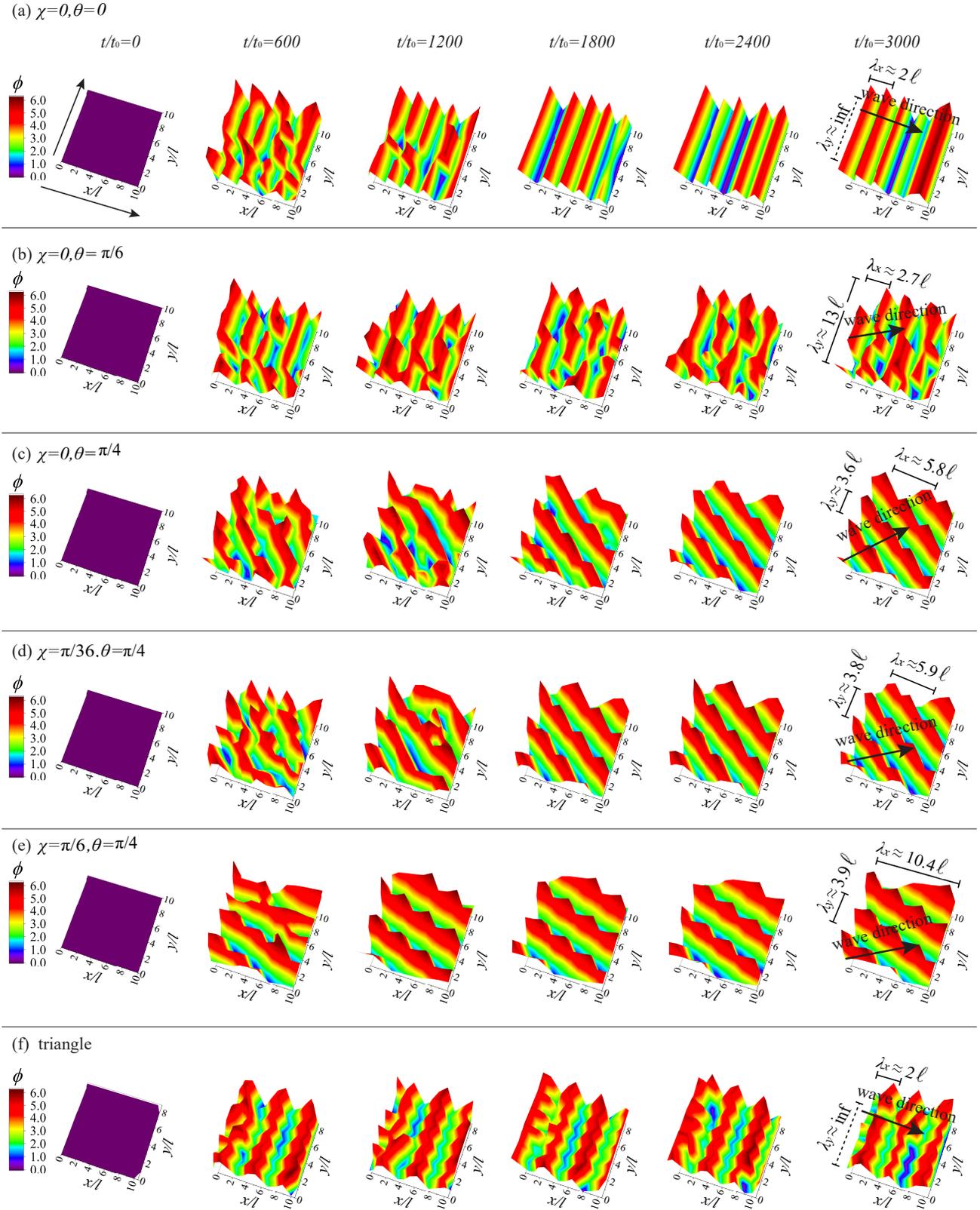}
\caption{Simulation snapshots showing how the phases that represent the cilia beating cycle evolve with time in an $11\times11$ array, where the cilia rotate along the trajectory defined by the angles: (a) $\chi=0$, $\theta=0$, (b) $\chi=0$, $\theta=\pi/6$, (c) $\chi=0$, $\theta=\pi/4$, (d) $\chi=\pi/36$, $\theta=\pi/4$ and (e) $\chi=\pi/6$, $\theta=\pi/4$ (see Fig. \ref{setup}). Panel (f) corresponds to a triangular lattice with $\chi=0$, $\theta=0$. Other parameters for panels (a)-(f) are: $a=0.2 \ell$,  $b=0.05\ell$, $h=\ell$, $A_{2}=0.5$, $B_{2}=0.5$, $C_{2}=0$, and $D_{2}=0$. The characteristic time scale is $t_{0}=\eta \ell^{2} /f_{0}$. The measured wavelengths and direction of propagation are shown on the last column; in every case the resulting wavevector lies within the range of the stable wave modes from the corresponding panels of Fig. \ref{relation}.}
\label{fig3}
\end{center}
\end{figure*}

\subsection*{Dispersion relation}
Let us now consider a situation where the cilia beat in coordination and generate a metachronal wave of frequency $\omega$ and wavevector $\bm{k}$. We describe the traveling wave as $\bar{\phi}(\bm{r}, t)=\omega t-\bm{k}\cdot\bm{r}+\delta\bar{\phi}_{\bm{k}}(\bm{r}, t)$,
where $\delta\bar{\phi}_{\bm{k}}$ represents perturbations around the harmonic traveling wave {\em ansatz}. Evaluating Eq. (\ref{m2}) at the zeroth order, and making use of the identity $ \sum_{\bm{r}'} e^{i\bm{q}\cdot(\bm{r}-\bm{r}')}= \sum_{\bm{G}}\frac{4\pi^{2}}{\ell^{2}}\delta^{2}(\bm{q}+\bm{G})$ where $\bm{G}$ represents the reciprocal lattice vectors, we find the dispersion relation of the metachronal waves as
\begin{equation}\label{omega5}
\omega(\bm{k})=\Omega_0 \left[1 +\frac{\pi}{4}\frac{h b}{\ell^{2}}\sum_{\bm{G}} {\cal M}(\bm{k}+\bm{G})\right].
\end{equation}
We recall, for example, that for a square lattice we have $\bm{G}=\frac{2\pi}{\ell}(m\bm{e}_{x}+n\bm{e}_{y})$ for $m,n\in \mathbb{Z}$.

The dispersion relation is plotted in Fig.~\ref{relation} for various choices of cilia orientation. Note that the resulting frequencies for all modes are somewhat larger than the single cilium frequency $\Omega_0$, due to a renormalization of the frequency by hydrodynamics interactions. For example, $\omega(\bm{0})/\Omega_0=1 +\frac{\pi}{4}\frac{h b}{\ell^{2}}\sum_{\bm{G}} {\cal M}(\bm{G}) \approx 1.43$ for the parameter set corresponding to Fig~\ref{relation}(a), namely, $\theta=0$, $\chi=0$, $h=\ell$, $a=0.2 \ell$, $b= 0.05 \ell$, $A_{2}=0.5$, $B_{2} =0.5$, $C_{2}=0$, and $D_{2} =0$. 

\subsection*{Stability criterion}
Satisfying the dispersion relation provides a necessary condition for frequencies and wavevectors to represent metachronal waves. However, it does not guarantee that the wave is a stable solution to Eq.~(\ref{m2}). To check the stability of a solution, we can expand Eq.~(\ref{m2}) in terms of $\delta\bar{\phi}_{\bm{k}}$, and probe the first order governing equation for the perturbation. In Fourier space, we find the time evolution of a perturbation with wavevector $\bm{q}$ in a background of uniform wave with wavevector $\bm{k}$ to satisfy the following equation
\begin{equation}\label{stability3}
\partial_{t}\delta \bar{\phi}_{\bm{k}}(\bm{q})=-\Big[\Gamma(\bm{q},\bm{k})-\Gamma(\bm{0},\bm{k})\Big] \delta\bar{\phi}_{\bm{k}}(\bm{q}),
\end{equation}
where
\begin{equation}\label{gamma1}
\Gamma(\bm{q},\bm{k})=\frac{\pi b \Omega_{0}}{8 \ell^{2}}\sum_{\bm{G}}\big[{\cal S}(\bm{q}+\bm{k}-\bm{G})+{\cal S}(\bm{q}-\bm{k}-\bm{G})\big].
\end{equation}
The sign of $\Gamma(\bm{q},\bm{k})-\Gamma(\bm{0},\bm{k})$ determines whether the background wave solution with wavevector $\bm{k}$ is stable with respect to a perturbation with wavevector $\bm{q}$. If $\Gamma(\bm{q},\bm{k})-\Gamma(\bm{0},\bm{k}) >0$ for all values of $\bm{q}$, then the background metachronal wave with wavevector $\bm{k}$ is linearly stable.

In Fig.~\ref{relation}, the wavevectors corresponding to linearly stable solutions of Eq.~(\ref{m2}) are shown as (blue) dots. The explicit expression for ${\cal S}$ allows us to make predictions about the necessary criteria for the stability of the waves. For example, when $\chi=0$ stability requires the condition $2 B_{2}-D_{2}>0$ to be satisfied. In this case, one can generally observe that the stable modes propagate along the direction of the ciliary beating with wavelengths (denoted by $\lambda$) that are in the range of $\lambda \gtrsim 2\ell$. One can observe that for directions that do not coincide with the lattice axes, the domains of permissible wavevectors shrink in size and tend towards larger wavelengths (see Figs. \ref{relation}(a)-(c)). Increasing the angle $\chi$, which amounts to tilting the ciliary beating orbit away from the $z$-axis, further accentuates this feature while allowing for the direction of propagation to deviate from the direction of beating, leading to the formation of dexioplectic or laeoplectic metachronism [see Figs. \ref{relation}(d)-(e)].

To examine the role of the underlying lattice structure, we consider a triangular lattice, 
which is characterized by reciprocal lattice vectors $\bm{G}=\frac{2\pi}{\ell}[(m\bm{e}_{x}+(-m/\sqrt{3}+2n/\sqrt{3}\bm{e}_{y})],~(m,n\in \mathbb{Z})$. The dispersion relation does not differ appreciably from that of a square lattice, as Fig.~\ref{relation}(f) shows for the case of $\theta=0$ and $\chi=0$. However, while for a square lattice the stable wave region is centered around $k_{x}=\pm\pi/\ell$ and $k_{y}=0$, for a triangular lattice it is centered around $k_{x}=\pm\pi/\ell$ and $k_{y}=0$ or $k_{x}=\pm(2-\sqrt{3})\pi/\ell$ and $k_{y}=\pi/\ell$, which means that there will be a phase shift between the neighboring cilia along $\bm{e}_{x}$ direction, hence giving rise to dexioplectic or laeoplectic waves \cite{Knight-Jones1954}. This behavior can be understood by analyzing the dynamics of two cilia with the appropriate geometric arrangement \cite{suppl}. We thus find that tuning the orientation of the cilia trajectories and controlling the positioning of the cilia in the array provide the possibility to generate metachronal waves with desired wavelengths and directions of propagation.

We note that in our current formulation the stability criterion is degenerate with respect to the direction of propagation, i.e. $\pm\bm{k}$ are both either stable or unstable at the same time. The symmetry can be broken by considering near-field effects in the hydrodynamic interaction between cilia; this will be discussed in future work. Note also that while the stability analysis is performed in terms of $\bar{\phi}$, the one-to-one correspondence in Eq.~(\ref{relationbetweenphase}) guarantees that it will also describe the stability of the modes in terms of the original $\phi$ coordinate.

\subsection*{Agent-based simulation}
To support the validity of the above analytical description, which is analyzed within the framework of linear stability analysis, we perform numerical simulations based on the governing dynamical equations of the cilia [Eq.~(\ref{Mo1})]. The examples of the time evolution are presented in Fig.~\ref{fig3} for an $11\times11$ cilia array, which is simulated with periodic boundary conditions. The cilia are positioned on a square lattice [Fig.~\ref{fig3}(a-e)] with different tilting angles $\theta$ and $\chi$, as well as a triangular lattice with $\theta=0$ and $\chi=0$ [Fig.~\ref{fig3}(f)]. The cilia are all initiated with the same phase $\phi=0$ at the start of the simulation at $t=0$. The time interval for each simulation step is $dt/t_{0}=2\cdot 10^{-3}$, with $t_{0}=\eta \ell^{2} /f_{0}$ defining a characteristic time.

As can be seen in Fig.~\ref{fig3} and the Supplemental Movies \cite{suppl}, in all cases the cilia coordinate with each other and form a metachronal wave after a transient period. For example, in the case of the cilia on a square lattice with the tilting angles of the trajectory as $\theta=0$ and $\chi=0$ [as shown in Fig.~\ref{fig3}(a)], the cilia beat in the form of the metachronal wave with the wave vector $k_{x}\simeq\pi/\ell$ and $k_{y}\simeq 0$. In another case, corresponding to $\theta=\pi/4$ and $\chi=0$, the cilia beat in the form of the metachronal wave with the wavevector $k_{x}\simeq pi/2\ell$ and $k_{y} \simeq \pi/2\ell$. These simulation results agree very well with the prediction of the linear stability analysis; the values of the measured stable waves lie for all cases within the range predicted by the theoretical calculations as presented in Fig.~\ref{relation}.

\section*{Discussion}
We have constructed a theoretical framework to study metachronal waves in ciliary arrays, where each cilium is driven independently with the same beat pattern and interacts with the others via hydrodynamic interactions for arbitrary geometric configurations. We calculate the dispersion relation of the system, relating the propagation frequency and the wavevector of the metachronal wave, and observe that the frequency is relatively insensitive to the changes in the wavevector. We have found that stable waves correspond to finite domains of wavevector, which are selected with relatively well-defined orientation of propagation that is determined by the geometric characteristics of the ciliary beating pattern and the lattice structure. Our results allow us to predict the role of the different harmonics in the moment decomposition of the beat pattern and the friction, which in turn can be used to make predictions about control of metachronal waves using external cues, as has been demonstrated in the case of phototaxis of {\em Chlamydomonas} \cite{Bennett2015}.

\appendix 

\section*{Synchronization of two cilia}
Consider two cilia rotating in $yz$ plane, and the centre of one cilium trajectory is located at $\bm{R}_{1}=(0, 0, h)$ with phase $\phi_{1}$, and the other one is located at $\bm{R}_{2}=(\ell\cos\Theta, \ell\sin\Theta, h)$ with phase $\phi_{2}$.
The dynamic equation of cilia $1$ is,
\begin{eqnarray}\label{Mo3_1}
 \dot{\phi}_{1} = \frac{f(\phi_{1})}{\zeta(\phi_{1}) a}+\frac{1}{a}
 \bm{t}_{1}\cdot \bm{G}(\bm{R}_{1};\bm{R}_{2})\cdot\bm{t}_{2} f(\phi_{2}),
\end{eqnarray}
or alternatively,
\begin{eqnarray}\label{Mo3_2}
 \dot{\phi}_{1} = \frac{f(\phi_{1})}{\zeta_{0} a}+ H_{12}\frac{f(\phi_{2})}{\zeta_{0} a},
\end{eqnarray}
with $H_{12}=\bm{t}_{1}\cdot \bm{G}(\bm{R}_{1};\bm{R}_{2})\cdot\bm{t}_{2} \zeta_{0}$.
After the coordinate transformation introduced in the main text, $\phi\rightarrow\bar{\phi}$, the dynamic equation of cilia $1$ can be re-written as:
\begin{eqnarray}\label{Mo3_3}
 \dot{\bar{\phi}}_{1} = \Omega_{0} \left[1 + \bar{H}_{12}(\bar{\phi}_{1},\bar{\phi}_{2})
 \frac{\bar{f}(\bar{\phi}_{2})}{\bar{f}(\bar{\phi}_{1})}\right],
\end{eqnarray}
where $\bar{H}_{12}(\bar{\phi}_{1},\bar{\phi}_{2})=H_{12}(\phi_{1},\phi_{2})$ and
$\bar{f}(\bar{\phi}_{1,2})=f(\phi_{1,2})$. $f(\phi) = f_{0}[1+A_{2}\cos 2\phi+B_{2}\sin 2\phi]$ is taken.
Then we can obtain
\begin{eqnarray}\label{Mo3_3}
 \dot{\bar{\Delta}} &=& \Omega_{0} \left[\bar{H}_{12}(\bar{\phi}_{1},\bar{\phi}_{2})
 \frac{\bar{f}(\bar{\phi}_{2})}{\bar{f}(\bar{\phi}_{1})}-\bar{H}_{2 1}(\bar{\phi}_{2},\bar{\phi}_{1})
 \frac{\bar{f}(\bar{\phi}_{1})}{\bar{f}(\bar{\phi}_{2})}\right]\nonumber\\
 &=& \Omega_{0} \bar{H}_{12}(\bar{\phi}_{1},\bar{\phi}_{2})\left[
 \frac{\bar{f}(\bar{\phi}_{2})}{\bar{f}(\bar{\phi}_{1})}-
 \frac{\bar{f}(\bar{\phi}_{1})}{\bar{f}(\bar{\phi}_{2})}\right],
\end{eqnarray}
where $\bar{\Delta}=\bar{\phi}_{1}-\bar{\phi}_{2}$.

In the case of $h\geq \ell \gg a$,
the Blake tensor for the cilia on a lattice coordinated as $\bm{r}$ can be approximated as:
\begin{eqnarray}\label{Blake3}
G_{\alpha\beta}&\simeq&\frac{1}{16\pi^{2}\eta}\int d^{2}q\frac{1}{|\bm{q}|}e^{i\bm{q}\cdot(\bm{r}'-\bm{r})}\left(2\delta_{\alpha\beta}-
\frac{q_{\alpha}q_{\beta}}{|\bm{q}|^{2}}\right),~(\alpha,\beta=x,y),\nonumber\\ \\
G_{\alpha z}&\simeq& 0,\\
G_{z\alpha}&\simeq& 0,\\
G_{zz}&\simeq&\frac{1}{16\pi^{2}\eta}\int d^{2}q\frac{1}{|\bm{q}|}e^{i\bm{q}\cdot(\bm{r}'-\bm{r})}.
\end{eqnarray}
where we ignore the difference between the projected positions $\bm{R}_{p}$ and the lattice locations $\bm{r}$, as well as fast decaying terms such as $\exp(-|\bm{q}|h)$.
By taking the Green's function in Equation \eqref{Blake3},  $\bar{H}_{12}(\bar{\phi}_{1},\bar{\phi}_{2})$ can be approximated as
$\bar{H}_{12}(\bar{\phi}_{1},\bar{\phi}_{2})=G_{yy}\sin\bar{\phi}_{1}\sin\bar{\phi}_{2} + G_{zz}\cos\bar{\phi}_{1}\cos\bar{\phi}_{2}$.
By taking $\bar{\sum}=\bar{\phi}_{1}+\bar{\phi}_{2}$ and averaging over the fast variable $\bar{\sum}$ in terms of $\langle ... \rangle = \frac{1}{4\pi}\int_{0}^{4\pi}...$,
then the dynamic equation can be written as
\begin{eqnarray}\label{Mo3_3}
 \dot{\bar{\Delta}} \simeq (G_{yy} - G_{zz}) B_{2} \sin \bar{\Delta}=
 \frac{\sin^{2}\Theta}{\ell}B_{2} \sin \bar{\Delta}.
\end{eqnarray}
Here we introduce an effective potential, which is
\begin{eqnarray}\label{Mo3_3}
 \mathcal{U} &=& -\int_{0} ^{\bar{\Delta}} d \bar{\Delta}'
 \frac{\sin^{2}\Theta}{\ell}B_{2} \sin \bar{\Delta}' \nonumber\\
 &=& \frac{\sin^{2}\Theta}{\ell}B_{2} \cos (\bar{\Delta} -1 )\simeq \frac{\sin^{2}\Theta}{\ell}B_{2} (\cos\Delta -1 ),
\end{eqnarray}
where the local minimum locates at $\Delta=\pi$ as the stable phase shift between the cilia.
By assuming the cilia beat with a wave vector $\bm{k}$,
then the phase difference is $\Delta=\bm{k} \cdot \bm{r} = k_{x}\ell\cos\Theta + k_{y}\ell\sin\Theta$, so the wave vector should follow $k_{x}\ell\cos\Theta + k_{y}\ell\sin\Theta = \pi$.

We are grateful to Andrej Vilfan and Masao Doi for fruitful discussions. This work was supported by the Max-Planck-Gesellschaft. F. M. thanks partial supports from Alexander von Humboldt Foundation, Strategic Priority Research Program of Chinese Academy of Sciences (No. XDA17010504) and the National Natural Science Foundation of China (No. 12047503). R. R. B. Acknowledges a doctoral scholarship from the EPSRC and a University of Bristol Vice-Chancellor's Fellowship.

\bibliography{refer}

\end{document}